# Determination of the magnetic permeability, electrical conductivity, and thickness of ferrite metallic plates using a multi-frequency electromagnetic sensing system

Mingyang Lu, Yuedong Xie, Wenqian Zhu, Anthony Peyton, and Wuliang Yin, *Senior Member, IEEE*

*Abstract*—In this paper, an inverse method was developed which can, in principle, reconstruct arbitrary permeability, conductivity, thickness, and lift-off with a multi-frequency electromagnetic sensor from inductance spectroscopic measurements.

Both the finite element method and the Dodd & Deeds formulation are used to solve the forward problem during the inversion process. For the inverse solution, a modified Newton–Raphson method was used to adjust each set of parameters (permeability, conductivity, thickness, and lift-off) to fit inductances (measured or simulated) in a least-squared sense because of its known convergence properties. The approximate Jacobian matrix (sensitivity matrix) for each set of the parameter is obtained by the perturbation method. Results from an industrial-scale multi-frequency sensor are presented including the effects of noise. The results are verified with measurements and simulations of selected cases.

The findings are significant because they show for the first time that the inductance spectra can be inverted in practice to determine the key values (permeability, conductivity, thickness, and lift-off) with a relative error of less than 5% during the thermal processing of metallic plates.

*Index Terms*—Electrical conductivity, electromagnetic sensor, inversion, lift-off, magnetic permeability, measurements, multi-frequency, non-destructive testing (NDT), thickness

## I. INTRODUCTION

MULTI-frequency electromagnetic sensors, such as EM-spec [1], are now being used to non-destructively test the properties of strip steel on-line during industrial processing. These sensors measure the relative permeability of the strip during process operations such as controlled cooling and the permeability values are analyzed in real time to determine important microstructural parameters such as the transformed fraction of the required steel phases. These parameters are critical to achieving the desired mechanical properties in the strip product. The inductance spectra produced by the sensor are not only dependent on the magnetic permeability of the strip but is also an unwanted function of the electrical conductivity and thickness of the strip and the distance between the strip steel and the sensor (lift-off). The confounding cross-sensitivities to these parameters need to be rejected by the processing algorithms applied to inductance spectra.

In recent years, the eddy current technique (ECT) [2-5] and the alternating current potential drop (ACPD) technique [6-8] were the two primary electromagnetic non-destructive testing techniques (NDT) [9-21] on metals' permeability measurements. However, the measurement of permeability is still a challenge due to the influence of conductivity, lift-off, and thickness of the detected signal. Therefore, decoupling the impact of the other parameters on permeability is quite vital in permeability measurement [22-24]. Some studies have been proposed for the ferrous metallic permeability prediction based on both the eddy current technique and alternative current potential drop method. However, these methods all use a low excitation frequency (typically 1 Hz-50 Hz), which may reduce the precision of the measurement. Yu has proposed a permeability measurement device based on the conductivity invariance phenomenon (CIP) [25], and the measured results tested by the device were proved to be accurate. The only imperfection of this device is requiring substrate metal on the top and bottom sides of the sample, which is impractical in some applications, for example, in cases where only one side of the sample is accessible. Adewale and Tian have proposed a design of novel PEC probe which would potentially decouple the influence of permeability and conductivity in Pulsed Eddy-Current Measurements (PEC) [26]. They reveal that conductivity effects are prominent on the rising edge of the transient response, while permeability effects dominate in the stable phase of the transient response; this is as we encountered in multi-frequency testing, as the rising edge of the transient response contains high-frequency components while the stable phase contains lower frequency components and low frequency is more related to permeability contribution due to magnetization. They use normalization to separate these effects.

This paper considers the cross-sensitivity of the complex spectra from a multi-frequency inductance spectrum to the four variables namely, permeability, conductivity, thickness, and lift-off with tested sensors. The paper then goes further to consider the solution of the inverse problem of determining unique values for the four variables from the spectra. There are two major computational problems in the reconstruction

This work was supported by [UK Engineering and Physical Sciences Research Council (EPSRC)] [grant number: EP/M020835/1] [title: Electromagnetic tensor imaging for in-process welding inspection]]. Paper no. TII-18-2870. (Corresponding author: Yuedong Xie.)

The authors are with the School of Electrical and Electronic Engineering, University of Manchester, Manchester, M13 9PL UK (e-mail: mingyang.lu@manchester.ac.uk; yuedong.xie@warwick.ac.uk; wenqian.zhu@manchester.ac.uk; a.peyton@manchester.ac.uk; wuliang.yin@manchester.ac.uk).

process: the forward problem and the inverse problem. The forward problem is to calculate the frequency-dependent inductance for metallic plates with arbitrary values of permeability, conductivity, thickness, and lift-off (i.e. the distance between the sensor and test sample). The inverse problem is to determine each profile's sensitivity, i.e. the changes in each profile (permeability, conductivity, thickness, and lift-off with tested sensors) from the changes in frequency-dependent inductance measurements. A dynamic rank method was proposed to eliminate the ill-conditioning of the problem in the process of reconstruction. Profiles of permeability, conductivity, thickness, and lift-off have been reconstructed from simulated and measured data using an EM sensor, which has verified this method.

## II. SAMPLES & FORWARD PROBLEM

Both the finite-element method and the Dodd and Deeds formulation [27] are used to solve the forward problem during the inversion process. The sensor is composed of three coaxially arranged coils, configured as an axial gradiometer; with the three coils having the same diameter. The central coil is a transmitter and the two outer coils are receivers and connected in series opposition. A photograph of the sensor is shown in figure 1, with its dimensions in Table I. The design of this sensor is such that both the measurements and the analytical solution of Dodd and Deeds are accessible, however, the geometry of the sensor has also been designed so that a high-temperature version can be fabricated for use at high temperatures in a production furnace and consequently magnetic components such as a magnetic yoke cannot be used. The detailed design of the industrial high-temperature version of the sensor is beyond the scope of this paper.

The samples were chosen to be a series of dual-phase steel (DP steel) samples - DP600 steel (with an electrical conductivity of 4.13 MS/m, relative permeability of 222, and thickness of 1.40 mm), DP800 steel (with an electrical conductivity of 3.81 MS/m, relative permeability of 144, and thickness of 1.70 mm), and DP1000 steel (with an electrical conductivity of 3.80 MS/m, relative permeability of 122, and thickness of 1.23 mm) and same planar dimensions of 500 × 400 mm size. The same probe was used for measurements at several lift-offs of 5 mm, 30 mm, 50 mm, and 100 mm. All these samples parameters are obtained from our previous work in [33]. The steels contained 0.1-0.2 wt% C and 1.5 – 2.2 wt% Mn, the amount of these elements generally increasing with increasing strength. Additions of Nb and Ti are also used to achieve strength levels of DP800 and DP1000. The exact chemical composition is confidential. The microstructure is produced by controlling the transformation of austenite after hot rolling. Metallographic samples were taken in the transverse direction, prepared to a 1/4-micron polish finish, and etched in 2% nital. The samples were imaged using a JEOL7000 SEM (SEM micrograph in figure 1 (c)). The ferrite, bainite/tempered martensite, and martensite phases were manually distinguished based on the contrast within the grains, and the percentage of each phase present was quantified using "Image J" image analysis software. Results are included in Table II.

For the experimental setup, a symmetric electromagnetic sensor was designed for steel micro-structure monitoring in the Continuous Annealing & Processing Line (CAPL). There are three coils winded for the CAPL sensor. The excitation coil sits in the middle and two receive coils at bottom and top respectively. One receive coil is used as the test coils; the other is used as a reference. The difference between the two receive coils is recorded. In order to better understand the CAPL sensor performance, a dummy sensor has been built for the lab use, shown in figure 1(a). The diameter of the sensor is 150 mm. Each of the coils has 15 turns, and the coil separation is 35 mm. Details of sensor dimensions are shown in Table I. Solartron Impedance Analyzer SI1260 is used to record the experimental sensor output data.

Steel users are placing increasing competitive pressure on producers to supply ever more sophisticated steel grades to tougher specifications, especially in the automobile and pipeline sectors. This drives the need to monitor microstructure online and in real time to help control material properties and guarantee product uniformity. To achieve this task, robust and process-compliant instrumentation is required. There are a small number of commercial systems that can assess steel quality by exploiting changes in magnetic properties. These systems typically operate at positions in the processing route where the steel is at ambient or relatively low temperatures. However, it is important to log and control microstructure during hot processing, where the hot steel is undergoing a dynamic transformation. Figure 1(b) shows the development and implementation of a new electromagnetic (EM) inspection system - EMspec for assessing microstructure during controlled cooling on a hot strip mill. The EM inspection system exploits magnetic induction spectroscopy, i.e., the frequency dependent response of the strip, to determine a transformation index which can characterize the evolution of the microstructure during cooling. This system is able to link microstructure of steel, via its EM properties, to the response of the EM inspection system overall.

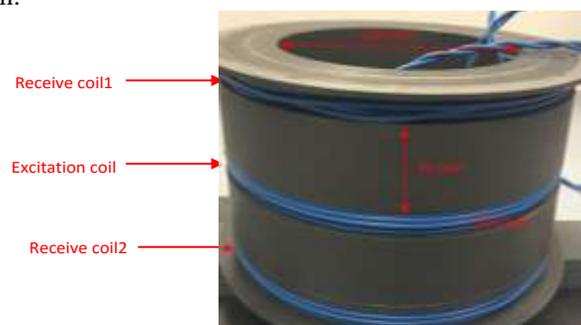

(a)

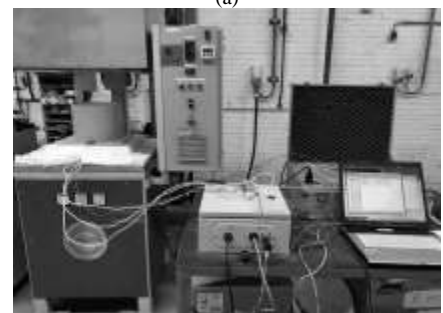

(b)

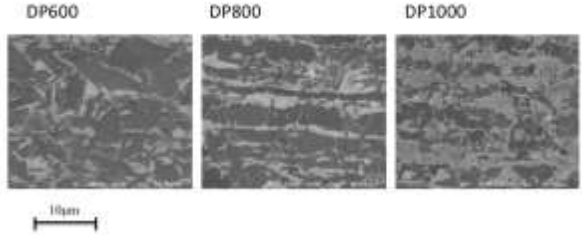

Fig. 1. (a)Sensor configuration (b) EMspec system (c) Images by JEOL7000 SEM micrograph

TABLE I
COILS PARAMETERS

| Parameters | Value | Unit |
|---|---|---|
| Inner diameter | 150 | mm |
| Outer diameter | 175 | mm |
| lift-offs | 5, 30, 50, 100 | mm |
| Coils height | 10 | mm |
| Coils gap | 35 | mm |
| Number of turns N1(Excitation coil) = N2(Receive coil1) = N3(Receive coil2) | 15 | / |

TABLE II
FERRITE FRACTIONS OF DP SAMPLES

| DP samples | Percentage of ferrite (%) |
|---|---|
| DP600 | 83.6 |
| DP800 | 78.4 |
| DP1000 | 40.0 |

Here, the Dodd Deeds analytical solution is chosen to be the forward problem solver.

The Dodd Deeds analytical solution describes the inductance change of an air-core coil caused by a layer of the metallic plate for both non-magnetic and magnetic cases [28, 29]. Another similar formula exists [30]. The difference in the complex inductance is $\Delta L(\omega) = L(\omega) - L_A(\omega)$, where the coil inductance above a plate is $L(\omega)$, and $L_A(\omega)$ is the inductance in free space.

The formulas of Dodd and Deeds are:

$$\Delta L(\omega) = K \int_0^\infty \frac{P^2(\alpha)}{\alpha^6} A(\alpha)\phi(\alpha)d\alpha \quad (1)$$

Where,

$$\phi(\alpha) = \frac{(\alpha_1 + \mu\alpha)(\alpha_1 - \mu\alpha) - (\alpha_1 + \mu\alpha)(\alpha_1 - \mu\alpha)e^{2\alpha_1 c}}{-(\alpha_1 - \mu\alpha)(\alpha_1 - \mu\alpha) + (\alpha_1 + \mu\alpha)(\alpha_1 + \mu\alpha)e^{2\alpha_1 c}} \quad (2)$$

$$\alpha_1 = \sqrt{\alpha^2 + j\omega\sigma\mu_r\mu_0} \quad (3)$$

$$K = \frac{\pi\mu_0 N^2}{h^2(r_1 - r_2)^2} \quad (4)$$

$$P(\alpha) = \int_{\alpha r_1}^{\alpha r_2} xJ_1(x)dx \quad (5)$$

$$A(\alpha) = e^{-\alpha(2l_0 + h + g)}(e^{-2\alpha h} + 1) \quad (6)$$

Where, $\mu_0$ denotes the permeability of free space. $\mu_r$ denotes the relative permeability of plate. N denotes the number of turns in the coil; $r_1$ and $r_2$ denote the inner and outer radii of the coil; while $l_0$ and h denote the lift-off and the height of the coil, g denotes the gap between the exciting coil and receiver coil.

Here, both the finite-element method (FEM) and Dodd & Deeds simulations were computed on a ThinkStation P510 platform with Dual Intel Xeon E5-2600 v4 Processor, with 16G RAM. FEM was scripted and computed by Ansys Maxwell; Dodd & Deeds method was simulated on MATLAB.

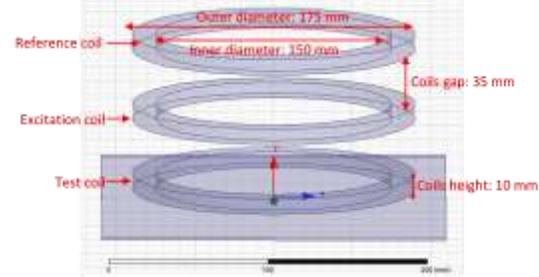

Fig. 2 Finite element modeling of the CAPL dummy sensor

FEM can also be used in this process. Ansys Maxwell is employed for the finite element modeling of the CAPL dummy sensor. The FEM 3D model is shown in figure 2 above. Both the sensor and steel sheet have the same dimension as the one used for the lab experiment.

For the experimental data, the real part of the inductance is defined from the mutual impedance of the transmitter and the receiver coils:

$$\mathrm{Im}(\Delta L) = \mathrm{Im}(\frac{Z(f) - Z_{air}(f)}{j2\pi f}) = \mathrm{Re}(\frac{-(Z(f) - Z_{air}(f))}{2\pi f}) \quad (7)$$

$$\mathrm{Re}(\Delta L) = \mathrm{Re}(\frac{Z(f) - Z_{air}(f)}{j2\pi f}) = \mathrm{Im}(\frac{Z(f) - Z_{air}(f)}{2\pi f}) \quad (8)$$

Where $Z(f)$ denotes the impedance of the coil with the presence of samples while $Z_{air}(f)$ is that of the coil in the air.

## III. INVERSE PROBLEM

The inverse problem, in this case, is to determine the permeability, conductivity, thickness, and lift-off with tested sensors profiles from the frequency-dependent inductance measurements. A modified Newton–Raphson method is used to adjust each profile to fit inductances (measured or simulated) in a least-squared sense because of its known convergence properties [35].

Definition of the problem is shown in follows.

1) $L_0 \in R^m$ : observed inductances arranged in a vector form (In this paper, a corresponded expansion matrix $\mathbf{L}_0$ with a real part and imaginary part of observed inductance listed on the top and bottom $m$ rows of the matrix - i.e. $\mathbf{L}_0 = [\mathrm{Re}(L_0); \mathrm{Im}(L_0)]$ is presented). And $m$ is the number of frequencies at which the inductance measurements are taken (here we select 10 frequency samples, i.e. $m = 10$).

2) $\sigma \in R$ : electrical conductivity of the tested sample.

3) $\mu \in R$ : permeability of the tested sample.

4) $t \in R$ : thickness of the tested sample.

5) $l \in R$ : lift-off of the sensors with respect to the sample plate.

6) $f : R^n \to R^m$ is a function mapping an input signal $[\sigma\ \mu\ t\ l]$ with $n$ degrees of freedom (here $n = 4$) into a set of $m$ approximate inductance observations(In this paper, a corresponded expansion matrix $\mathbf{f}$ with real part and imaginary part of observed inductance listed on the top and bottom $m$ rows of the matrix - i.e. $\mathbf{f} = [\text{Re}(f); \text{Im}(f)]$ is included). Here $\mathbf{f}$ can be calculated by the forward problem method such as Dodd and Deeds method.

7) $\phi = (1/2)[\mathbf{f} - \mathbf{L_0}]^T \cdot [\mathbf{f} - \mathbf{L_0}]$ is the squared error of the measured and estimated inductance.

Note that $\mathbf{f}$ is a function of sample's properties $(\sigma, \mu, t, l)$ under fixed measurement arrangements. The problem is to find a point $(\sigma*, \mu*, t*, l*)$ that is at least a local minimum of $\phi$. To find a candidate value of $(\sigma*, \mu*, t*, l*)$ that minimize $\phi$, $\phi$ is differentiated with respect to $(\sigma, \mu, t, l)$ and the result is set equal to the zero vector $\mathbf{0}$.

$$\phi' = [\mathbf{f}']^T[\mathbf{f} - \mathbf{L_0}] = \mathbf{0} \quad (9)$$

The term $\mathbf{f}'$ is known as the Jacobian matrix, an m×n matrix defined by (13).

Since $[\mathbf{f}']_{i,j}$ is still a nonlinear function of $(\sigma, \mu, t, l)$, the Taylor series expansion of $[\mathbf{f}']_{i,j}$ is taken from the reference point $(\sigma_r, \mu_r, t_r, l_r)$ and keeping the linear terms

$$\phi' \approx \phi'([\sigma_r, \mu_r, t_r, l_r]) + \phi''([\sigma_r, \mu_r, t_r, l_r])[\Delta\sigma \, \Delta\mu \, \Delta t \, \Delta l]^T \quad (10)$$

Where, $[\Delta\sigma \, \Delta\mu \, \Delta t \, \Delta l]^T = [\sigma \, \mu \, t \, l]^T - [\sigma_r \, \mu_r \, t_r \, l_r]^T$. The term $\phi''(\sigma_r, \mu_r, t_r, l_r)$ is called the Hessian matrix, which is difficult to calculate explicitly, but can be approximated within the small region about $[\sigma_r \, \mu_r \, t_r \, l_r]^T$ by

$$\phi''(\sigma_r, \mu_r, t_r, l_r) = [\mathbf{f}'(\sigma_r, \mu_r, t_r, l_r)]^T[\mathbf{f}'(\sigma_r, \mu_r, t_r, l_r)] \quad (11)$$

Substituting (9) and (11) into (10) and solving for $[\Delta\sigma \, \Delta\mu \, \Delta t \, \Delta l]^T$, we obtain

$$[\Delta\sigma \, \Delta\mu \, \Delta t \, \Delta l]^T = -\left[[\mathbf{f}'(\sigma_r, \mu_r, t_r, l_r)]^T \mathbf{f}'(\sigma_r, \mu_r, t_r, l_r)\right]^{-1} [\mathbf{f}'(\sigma_r, \mu_r, t_r, l_r)]^T [\mathbf{f}(\sigma_r, \mu_r, t_r, l_r) - \mathbf{L_0}] \quad (12)$$

Where, $\mathbf{f}(\sigma_r, \mu_r, t_r, l_r)$ is the calculated inductance for conductivity profile $(\sigma, \mu, t, l)$ using the forward solution, and $\mathbf{L_0}$ is the measured inductance for the sample. From (12), in order to calculate $[\Delta\sigma \, \Delta\mu \, \Delta t \, \Delta l]^T$, we need to have the sensitivity matrix $\mathbf{f}'(\sigma_r, \mu_r, t_r, l_r)$, which can be written in a matrix form $\mathbf{f}'(\sigma_r, \mu_r, t_r, l_r) = [\text{Re}(f'); \text{Im}(f')]$ with,

$$\mathbf{f}' = \begin{bmatrix} \frac{\partial f_1}{\partial \sigma}, \frac{\partial f_1}{\partial \mu}, \frac{\partial f_1}{\partial t}, \frac{\partial f_1}{\partial l} \\ \frac{\partial f_2}{\partial \sigma}, \frac{\partial f_2}{\partial \mu}, \frac{\partial f_2}{\partial t}, \frac{\partial f_2}{\partial l} \\ \vdots \\ \frac{\partial f_m}{\partial \sigma}, \frac{\partial f_m}{\partial \mu}, \frac{\partial f_m}{\partial t}, \frac{\partial f_m}{\partial l} \end{bmatrix} \quad (13)$$

One method of obtaining $\mathbf{f}'(\sigma_r, \mu_r, t_r, l_r)$ is to derive it from the Dodd and Deeds forward formulation (1). However, the resulting expression would be extremely complex even for more parameters needed to be estimated. Alternatively, the perturbation method can be used. The principle of the perturbation method is that the sensitivity of the inductance versus the $(\sigma, \mu, t, l)$ (essentially $\mathbf{f}'(\sigma_r, \mu_r, t_r, l_r)$) can be approximated by the inductance change, in response to a small perturbation from one of the $(\sigma, \mu, t, l)$, divided by the permeability change. Therefore, $\mathbf{f}'(\sigma_r, \mu_r, t_r, l_r)$ can be calculated in a column-wise fashion. The sensitivity matrix (14) can be obtained by dividing the inductance changes caused by a small parameter's change.

$$\begin{bmatrix} \frac{\partial f_1}{\partial \sigma}, \frac{\partial f_1}{\partial \mu}, \frac{\partial f_1}{\partial t}, \frac{\partial f_1}{\partial l} \\ \frac{\partial f_2}{\partial \sigma}, \frac{\partial f_2}{\partial \mu}, \frac{\partial f_2}{\partial t}, \frac{\partial f_2}{\partial l} \\ \vdots \\ \frac{\partial f_m}{\partial \sigma}, \frac{\partial f_m}{\partial \mu}, \frac{\partial f_m}{\partial t}, \frac{\partial f_m}{\partial l} \end{bmatrix} = \begin{bmatrix} \frac{\Delta I_{f_1}}{\Delta\sigma}, \frac{\Delta I_{f_1}}{\Delta\mu}, \frac{\Delta I_{f_1}}{\Delta t}, \frac{\Delta I_{f_1}}{\Delta l} \\ \frac{\Delta I_{f_2}}{\Delta\sigma}, \frac{\Delta I_{f_2}}{\Delta\mu}, \frac{\Delta I_{f_2}}{\Delta t}, \frac{\Delta I_{f_2}}{\Delta l} \\ \vdots \\ \frac{\Delta I_{f_m}}{\Delta\sigma}, \frac{\Delta I_{f_m}}{\Delta\mu}, \frac{\Delta I_{f_m}}{\Delta t}, \frac{\Delta I_{f_m}}{\Delta l} \end{bmatrix} \quad (14)$$

To use (14) for the calculation of the sensitivity is essentially a first-order finite difference approach to approximate the derivatives. To evaluate the effect of using finite changes of $(\sigma, \mu, t, l)$ in (14), different values of $(\Delta\sigma \, \Delta\mu \, \Delta t \, \Delta l)$ were used to calculate the sensitivity matrix. It is found that as we decrease the changes of $(\Delta\sigma \, \Delta\mu \, \Delta t \, \Delta l)$, the sensitivity map approach a set of slightly increased absolute values. However, as can be seen from figures 3-6, further decreasing $(\Delta\sigma \, \Delta\mu \, \Delta t \, \Delta l)$ would not make a significant difference to sensitivity.

The physical phenomena show that the eddy currents decay exponentially or diffuse from the surface into the metal. In its discrete form, the ill-conditioning in the Hessian matrix can result in the magnification of measurement error and numerical error in the reconstructed permeability profile. The singularity of the Hessian matrix is caused by the insensitivity or the mutual inductance with respect to one of the parameters under a specific frequency. For instance, the mutual inductance will be immune to the thickness on a specific high frequency due to the skin effect. Previously, the Tikhonov regularization method has been widely used in many inverse problems to deal with the ill-conditioning. However, the estimated error resulting from the regularization cannot be neglected due to the amendment of the sensitivity matrix. Here, a dynamic rank method is adopted to maintain that the results are estimated from the original unmodified sensitivity matrix, which has much improved the estimation accuracy. To simplify the notation, using $\mathbf{J}$ to represent $\mathbf{f}'(\sigma_r, \mu_r, t_r, l_r)$, (12) becomes

$$[\Delta\sigma \, \Delta\mu \, \Delta t \, \Delta l]^T = -\left[\mathbf{J}^T\mathbf{J}\right]^{-1}\mathbf{J}^T\left[\mathbf{f}(\sigma_r, \mu_r, t_r, l_r) - \mathbf{L_0}\right] \quad (15)$$

$$[\sigma \, \mu \, t \, l]^T = [\sigma_r \, \mu_r \, t_r \, l_r]^T + [\Delta\sigma \, \Delta\mu \, \Delta t \, \Delta l]^T \quad (16)$$

The principle of the dynamic rank method is indexing the columns whose elements are all zeros or nearly zeros (This because some parameters may not influence or sensitive to the inductance under a certain frequency, as shown in Fig.3, i.e. the conductivity sensitivity map). Then reduce the rank of the sensitivity matrix $\mathbf{J}$ by omitting the indexed columns. For each step in the iterative procedure, the corresponded rows of the estimated $[\Delta\sigma \, \Delta\mu \, \Delta t \, \Delta l]^T$ should be valued zeros.

Equations (15) and (16) can be used in an iterative fashion to find ($\sigma^*, \mu^*, t^*, l^*$). This formulation is known as the Gauss-Newton method. For each step in the iterative procedure, the Jacobian matrix **J** needs to be updated, which involves a considerable amount of computation.

## IV. PARAMETERS SENSITIVITY OF MULTI-FREQUENCY SPECTRA

The following figures illustrate the effects of different delta profiles ($\Delta\sigma\ \Delta\mu\ \Delta t\ \Delta l$) on both the real part (a) and imaginary part (b) of the sensor and samples mutual inductance change rate on the referred point ($\sigma_r\ \mu_r\ t_r\ l_r$) relative to samples' electrical conductivity ($\frac{Re(\Delta L)}{\Delta\sigma}$ & $\frac{Im(\Delta L)}{\Delta\sigma}$), relative permeability ($\frac{Re(\Delta L)}{\Delta\mu}$ & $\frac{Im(\Delta L)}{\Delta\mu}$), thickness ($\frac{Re(\Delta L)}{\Delta t}$ & $\frac{Im(\Delta L)}{\Delta t}$) and lift-off ($\frac{Re(\Delta L)}{\Delta l}$ & $\frac{Im(\Delta L)}{\Delta l}$). Here the referred point ($\sigma_r\ \mu_r\ t_r\ l_r$) is selected to be the properties of DP 600 steel sample with property profiles of (4.13 MS/m 222 1.4 mm 5 mm).

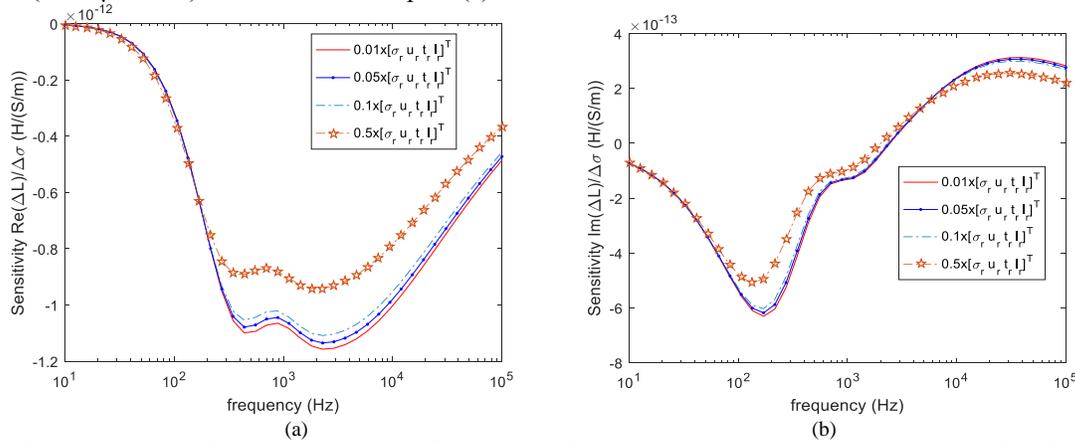

Fig. 3 Effects of different ($\Delta\sigma\ \Delta\mu\ \Delta t\ \Delta l$) on both real part (a) and imaginary part (b) of conductivity sensitivity of the referred point ($Re(\Delta L)/\Delta\sigma$ & $Im(\Delta L)/\Delta\sigma$)

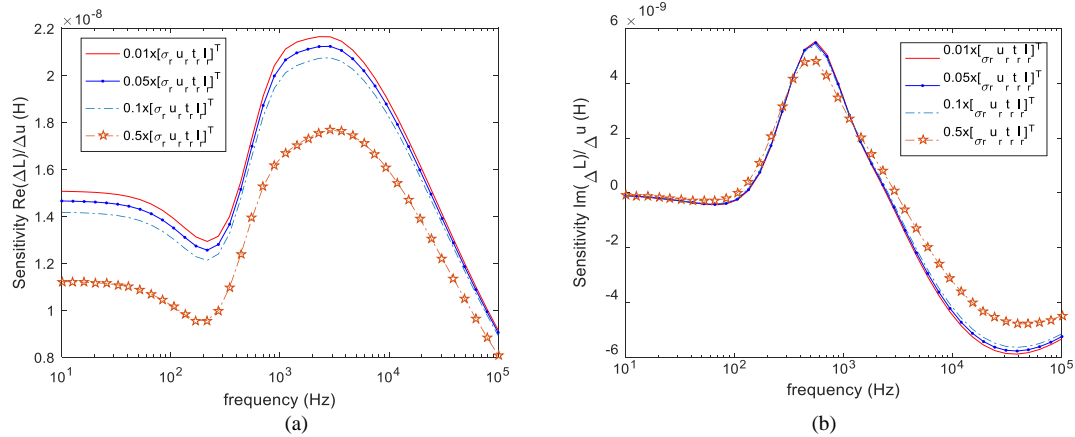

Fig. 4 Effects of different ($\Delta\sigma\ \Delta\mu\ \Delta t\ \Delta l$) on both real part (a) and imaginary part (b) of relative permeability sensitivity of the referred point ($Re(\Delta L)/\Delta\mu$ & $Im(\Delta L)/\Delta\mu$)

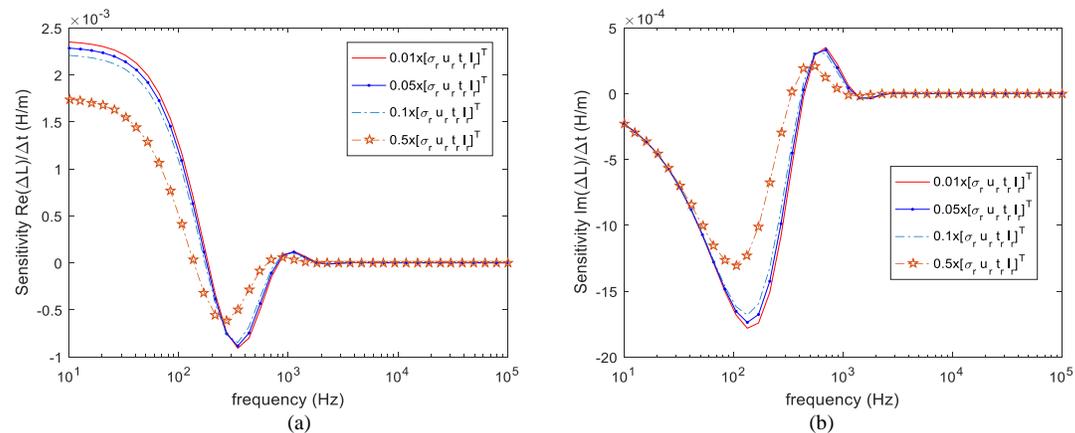

Fig. 5 Effects of different ($\Delta\sigma\ \Delta\mu\ \Delta t\ \Delta l$) on both real part (a) and imaginary part (b) of sample thickness sensitivity of the referred point ($Re(\Delta L)/\Delta t$ & $Im(\Delta L)/\Delta t$)

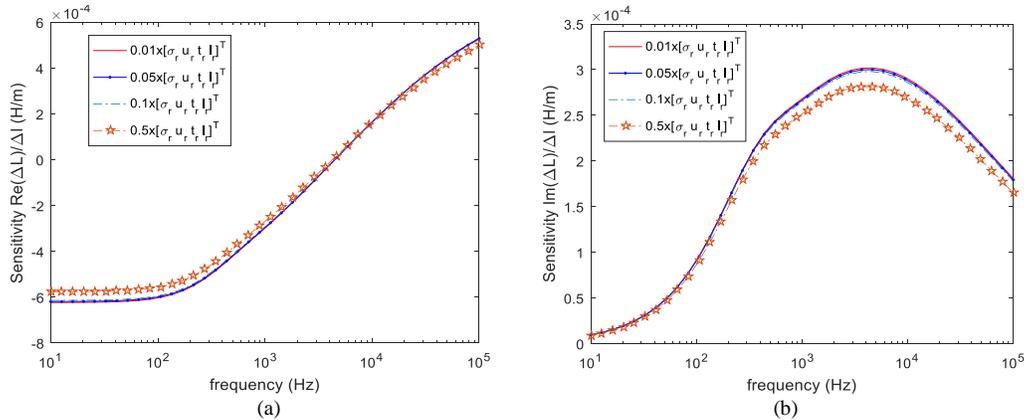

Fig. 6 Effects of different $(\Delta\sigma\ \Delta\mu\ \Delta t\ \Delta l)$ on both real part (a) and imaginary part (b) of sensors lift-offs sensitivity of the referred point $(Re(\Delta L)/\Delta l\ \&\ Im(\Delta L)/\Delta l)$

Figures 3 to 6 show the frequency-dependent sensitivity of the sample electrical conductivity, relative permeability, sample thickness, and sensor lift-off when different delta profiles $[\Delta\sigma\ \Delta\mu\ \Delta t\ \Delta l]^T$ within in the sensitivity matrix are selected to be 1%, 5%, 10% and 50% of referred properties $[\sigma_r\ \mu_r\ t_r\ l_r]^T$ ([4.13 MS/m 222 1.4 mm 5 mm]) respectively. It is found that as we decrease the changes of $(\Delta\sigma\ \Delta\mu\ \Delta t\ \Delta l)$, the sensitivity curves approach a set of saturation curves. Further decreasing $(\Delta\sigma\ \Delta\mu\ \Delta t\ \Delta l)$ would not make a significant effect on sensitivity spectra. Moreover, as can be seen from figure 3 to 6, the thickness sensitivity generally leads the parameters effect on inductance change rate. Since small changes in thickness will result in significant changes in the inductance when compared with other parameters, the reconstructed sample's thickness, in general, should be the most accurate values among the reconstructions of the samples' properties (electrical conductivity relative permeability μ, sample thickness t, and sensors lift-offs $l$).

## V. RECONSTRUCTION

As can be seen from Table III, the samples profiles are reconstructed more accurately from the Dodd and Deeds analytical solution with a relative error of less than 5%, which is achieved by utilizing the proposed dynamic rank method to eliminate the ill-conditioning problem in the process of reconstruction. Currently, there is still no commercial system that can simultaneously predict the four parameters (i.e. electrical conductivity, magnetic permeability, thickness, and lift-off) from the measured inductance/impedance signals. Commonly, most of the commercial system can accurately predict single parameter from the measurements. Here, the initial values $[\sigma_r\ \mu_r\ t_r\ l_r]^T$ for the iterative search of the solution are 5M S/m, 100, 2 mm, and 4 mm; The FEM method used is a custom-built solver software package which is more efficient than the canonical FEM method especially on the frequencies-sweeping mode. The solution of the field quantities under each frequency, which involves solving a system of linear equations using the conjugate gradients squared (CGS) method, is accelerated by using an optimized initial guess-the final solution from the previous frequency. More details of the custom-built FEM software package are included in [34]. The steel samples are finely meshed into a total number of 369 k elements prior to the FEM calculation. Besides, the inversion solver using Dodd and Deeds analytical method shows a more efficient performance than FEM due to a significantly reduced iteration number and operation time. Therefore, the following results are all deduced from the inversion method using Dodd and Deeds method.

TABLE III
RECONSTRUCTION OF THE SELECTED SAMPLES' PROPERTIES (ELECTRICAL CONDUCTIVITY RELATIVE PERMEABILITY, SAMPLE THICKNESS, SENSORS LIFT-OFFS) WHEN CALCULATED BY THE PROPOSED INVERSE SOLVER

| | Actual value | | | | | Estimated value by the proposed inverse solver using Dodd and Deeds | | | | | Estimated value by the proposed inverse solver using FEM | | | | |
|---|---|---|---|---|---|---|---|---|---|---|---|---|---|---|---|
| Case No. | 1 | 2 | 3 | 4 | 5 | 1 | 2 | 3 | 4 | 5 | 1 | 2 | 3 | 4 | 5 |
| Conductivity - σ (M S/m) | 4.13 | 3.81 | 3.80 | 3.80 | 3.80 | 4.06 | 3.70 | 3.68 | 3.65 | 3.63 | 4.03 | 3.67 | 3.65 | 3.63 | 3.59 |
| Predicted σ error (%) | / | / | / | / | / | 1.69 | 2.89 | 3.16 | 3.95 | 4.47 | 2.42 | 3.67 | 3.95 | 4.47 | 5.53 |
| Relative permeability - μ | 222 | 144 | 122 | 122 | 122 | 229 | 138 | 120 | 119 | 116 | 231 | 134 | 117 | 116 | 113 |
| Predicted μ error (%) | / | / | / | / | / | 3.15 | 4.17 | 1.64 | 2.46 | 4.92 | 4.05 | 6.94 | 4.10 | 4.92 | 7.38 |
| Thickness - t (mm) | 1.40 | 1.70 | 1.23 | 1.23 | 1.23 | 1.41 | 1.69 | 1.23 | 1.23 | 1.24 | 1.42 | 1.65 | 1.22 | 1.21 | 1.25 |
| Predicted t error (%) | / | / | / | / | / | 0.71 | 0.59 | 0 | 0 | 0.81 | 1.43 | 2.94 | 0.81 | 1.63 | 1.63 |
| Lift-off - $l$ (mm) | 5 | 5 | 5 | 30 | 50 | 5.02 | 5.03 | 5.06 | 30.41 | 50.63 | 5.04 | 5.05 | 5.08 | 30.83 | 50.92 |
| Predicted $l$ error (%) | / | / | / | / | / | 0.40 | 0.60 | 1.20 | 1.37 | 1.26 | 0.80 | 1.00 | 1.60 | 2.77 | 1.84 |
| Iteration No. | / | / | / | / | / | 7 | 5 | 4 | 15 | 22 | 89 | 67 | 103 | 77 | 92 |
| Computation time (seconds) | / | / | / | / | / | 21 | 19 | 23 | 17 | 26 | 556 | 523 | 583 | 537 | 563 |

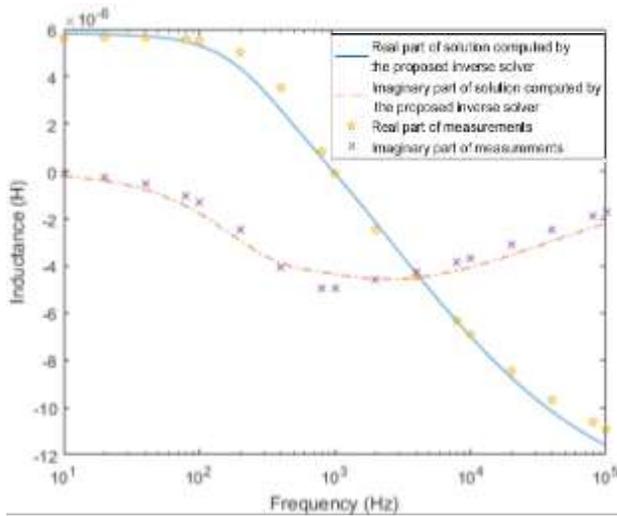

Fig. 7 Proposed inverse solver results and measurements of DP600 steel inductance multi-frequency spectra

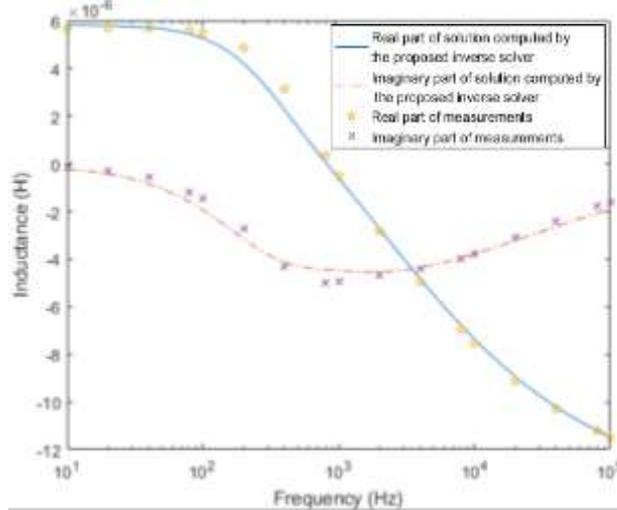

Fig. 8 Proposed inverse solver results and measurements of DP800 steel inductance multi-frequency spectra

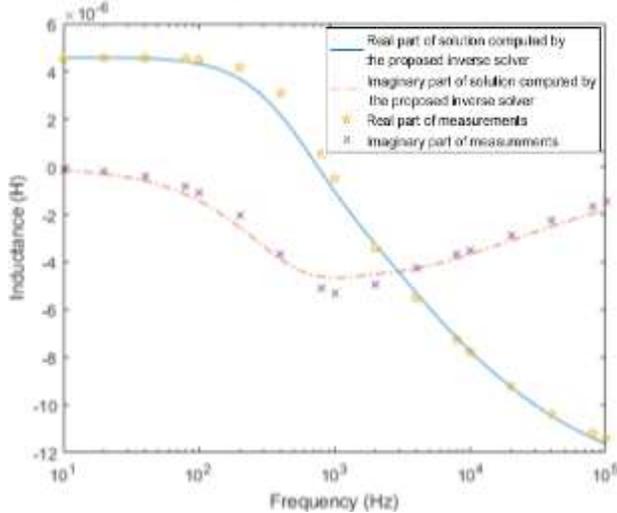

Fig. 9 Proposed inverse solver results and measurements of DP1000 steel inductance multi-frequency spectra

Figures 7 - 9 shows the inductance multi-frequency spectra of DP600, DP800, and DP1000 steel for both simulations from the estimated samples' properties calculated by the proposed inverse solver and measured results under a lift-off of 5 mm. It can be seen that the real part and imaginary part of the measured inductance multi-frequency spectra curves are close to that of the proposed inverse solver results for all the DP steel samples.

In practice, the observed inductance $L_0$ contains noise. Therefore, in this part, series of inductance $L_0$ are produced by adding noise to the observed inductance $L_0$. The noise has an amplitude value of 1%, 5% and 10% of $L_0$ and fluctuate randomly with frequency (i.e. $L_0 \pm 1\% \times L_0 \times R(f)$, $L_0 \pm 5\% \times L_0 \times R(f)$, $L_0 \pm 10\% \times L_0 \times R(f)$ with $R(f)$ randomly fluctuate in the range from 0 to 1 with frequencies). And the noise effect on the estimation of DP600 steel sample is illustrated in Table IV.

TABLE IV
NOISE EFFECT ON THE ESTIMATION OF DP600 STEEL SAMPLE PROPERTIES WHEN CALCULATED BY THE PROPOSED INVERSE

| Parameters | Actual value | Estimated value by the proposed inverse solver | | | | Unit |
|---|---|---|---|---|---|---|
| Fluctuate noise threshold (error magnitude) | / | 0 | 1% | 5% | 10% | / |
| Conductivity - σ | 4.13 | 4.06 | 4.03 | 4.27 | 4.43 | M S/m |
| Predicted σ error | / | 1.69 | 2.42 | 3.39 | 7.26 | % |
| Relative permeability - μ | 222 | 229 | 213 | 209 | 203 | / |
| Predicted μ error | / | 3.15 | 4.05 | 5.86 | 8.56 | % |
| Thickness - t | 1.40 | 1.41 | 1.41 | 1.42 | 1.45 | mm |
| Predicted t error | / | 0.71 | 0.71 | 1.43 | 3.57 | % |
| Lift-off - l | 5 | 5.02 | 4.96 | 4.93 | 4.84 | mm |
| Predicted l error | / | 0.40 | 0.80 | 1.40 | 3.20 | % |
| Iteration No. | / | 7 | 9 | 18 | 25 | / |

As can be seen from Table IV, with the introduction of measurements noise, the reconstructed parameters move further away from its actual value. But the reconstruction is still accurate with a relative error of less than 8.6%. Same trends have been observed for DP800 and DP1000 steel samples.

VI. CONCLUSION

In this paper, a method is presented which has the potential to reconstruct an arbitrary permeability, conductivity, thickness, and lift-off from inductance spectroscopic measurements with an EM sensor. The forward problem was solved numerically using both the finite-element method (FEM) and the Dodd and Deeds formulation [13].

Normally, the Dodd and Deeds analytical method is the primary choice, as it is much faster than FEM. For this reason, the proposed solver has its limitations – it requires lots of computation time for the reconstruction of the parameters for the samples excluded from the plate and cylinder geometry, such as a bent or defected plate. This is because the Dodd and Deeds methods can only valid for the simulation of plate and cylinder geometry.

In the inverse solution, a modified Newton–Raphson method was used to adjust the permeability profile to fit inductances (measured or simulated) in a least-squared sense. In addition, a dynamic rank method was proposed to eliminate the ill-conditioning of the problem in the process of reconstruction. Permeability, conductivity, thickness, and lift-off have been reconstructed from simulated and measured data with a small error of 5% only within an operation time 30seconds. However, the actual permeability used in our paper is only under room temperature. In fact, the steel's permeability will change with temperature; and the changes rate varies for different types of steel material, which will require lots of further measurements. Therefore, the inversion method performance of steels under different temperature should be analyzed for the next step.